# Combined Sum of Squares Penalties for Molecular Divergence Time Estimation

Peter J. Waddell[1]     Prasanth Kalakota[2]

waddell@med.sc.edu     kalakota@cse.sc.edu

[1] SCCC, University of South Carolina, Columbia, SC 29203, U.S.A.

[2] Department of Computer Science and Engineering, University of South Carolina, Columbia, SC  29208, U.S.A.

Estimates of molecular divergence times when rates of evolution vary require the assumption of a model of rate change. Brownian motion is one such model, and since rates cannot become negative, a log Brownian model seems appropriate. Divergence time estimates can then be made using weighted least squares penalties. As sequences become long, this approach effectively becomes equivalent to penalized likelihood or Bayesian approaches. Different forms of the least squares penalty are considered to take into account correlation due to shared ancestors. It is shown that a scale parameter is also needed since the sum of squares changes with the scale of time. Errors or uncertainty on fossil calibrations, may be folded in with errors due to the stochastic nature of Brownian motion and ancestral polymorphism, giving a total sum of squares to be minimized. Applying these methods to placental mammal data the estimated age of the root decreases from 125 to about 94 mybp. However, multiple fossil calibration points and relative molecular divergence times inflate the sum of squares more than expected. If fossil data are also bootstrapped, then the confidence interval for the root of placental mammals varies widely from ~70 to 130 mybp. Such a wide interval suggests that more and better fossil calibration data is needed and/or better models of rate evolution are needed and/or better molecular data are needed. Until these issues are thoroughly investigated, it is premature to declare either the old molecular dates frequently obtained (e.g. > 110 mybp) or the lack of identified placental fossils in the Cretaceous, more indicative of when crown-group placental mammals evolved.

## 1  Introduction

Weighted evolutionary trees, or phylogenies with edge (branch) lengths, can be used to estimate the relative ages of nodes, representing ancestors, in the tree. One of the first examples of this was the work of Wilson and Sarich inferring that humans and African apes diverged about 5 million years ago (Wilson and Sarich 1969). If the branch lengths of the tree are only slightly different from clock-like, then it is appealing to treat the evolutionary rate across the tree as the same (Hasegawa, Kishino, and Yano 1987). However, as the tree deviates from being clock-like, and as sequences become longer, it becomes apparent that it is important to estimate different evolutionary rates in different parts of the tree.

One way of estimating rates across the tree is to have a model of how rates change. A reasonable possibility is that evolutionary rates are inherited, which means that rates are autocorrelated (that is, correlated with rates at earlier times in that lineage). Another reasonable assumption is that the process of change undergoes Brownian motion, an assumption underlying many phylogenetic techniques for fitting quantitative characters onto a tree. Thorn and Kishino implemented such a model using a Bayesian technique (Thorne, Kishino, and Painter 1998). They used the additional assumption that it is the log evolutionary rates that undergo Brownian motion, a model often used in modeling other stochastic processes, such as the value of the stock market, which cannot take on negative values. Recently, Kitazoe et al. showed that with long sequences, the ML solution to estimating rates under the assumption of a log Brownian model, can be closely approximated with a weighted least squares approach (Kitazoe et al. 2007). They also point out that a popular least squares method proposed by Sanderson (1997) is not non-parametric, but rather based on the assumption that the rate of molecular evolution is a random walk with a jump of rate chosen from a normal distribution, every time a node is crossed. Another important connection made in Kitazoe et al. is that with long sequences, a least squares approach converges to the answers expected by either a Bayesian approach (Thorne, Kishino, and Painter 1998) or a penalized ML approach (Sanderson 2002), making this a powerful framework for better understanding properties of divergence time estimation.

A topic of particular interest is how old the root of the placental mammals are, which is tied up with the question of how old the superordinal crown groups are (Waddell, Okada and Hasegawa 1999). Whereas paleontologists favor an age fairly close to the KT boundary, for example 80 million years before present (mybp), to explain the lack of any identifiable placental fossils from the whole Cretaceous period (Alroy 1999), molecular dates cover a wide range and often tend to be in excess of 100 million years (Waddell et al. 1999, Waddell, Kishino, and Ota 2001, Springer et al. 2004). In order to move beyond this disagreement some of the issues that need to be addressed are comprehensive confidence intervals on molecular divergence times (Waddell and Penny 1996, Waddell et al. 1999), and a need to better understand the estimation and modeling of rate change (Thorne et al. 1998, Waddell, Kishino and Ota 2001, Kitazoe et al. 2007). Comprehensive errors on divergence times must treat fossil data not simply as a calibration point, nor as a set of arbitrary constraints, but as a distribution that needs to be folded in with all other sources of error including the error distribution generated by finite sequence lengths, ancestral polymorphism, and changing evolutionary rates (Waddell and Penny 1996).

Here we consider a revised penalty for Brownian evolution that takes into account correlation due to the ancestral branch and also considers the splitting of a Brownian process. Ways of treating rate change at the root are considered in the same framework. An extension of the methods in (Waddell et al. 1999) shows how fossil calibration errors may be folded in with errors due to the stochastic nature of the evolution of evolutionary rates and fluctuations caused by ancestral population sizes. A need to resample fossil data also is addressed using a bootstrap. The need for a scale factor (e.g., Waddell et al. 2007) for stabilizing least squares penalties of rate change is also examined. Methods are illustrated using data from Waddell et al. (2001) to reestimate placental divergence times.

## 2 Materials and Methods

The weighted tree of placental mammals (Appendix 1) is that used in Waddell et al. to estimate divergence times (Waddell, Kishino, and Ota 2001). This tree is based on a larger alignment (that of figure 2b of their work) than other studies looking at mammal divergence times and includes the coding sequences of both nuclear and mitochondrial protein genes. Indeed, nearly all later articles looking at deep mammalian divergence times, including examples such as Springer et al. (2003) and Kitazoe et al. (2007), use what are essentially subsets of this data set. Larger alignments reduce the impact of stochastic errors on edge length estimates which is important, since as these errors tend to zero they can be safely ignored from both a Bayesian and a penalized likelihood perspective (Kitazoe et al. 2007). The model of edge length estimation (JTT, $\Gamma$) remains that favored in current work (Kitazoe et al. 2007).

Minimization of penalty functions was done using a numerical optimization package (Frontline Systems 2004) that includes a generalized reduced gradient method (Lasdon et al. 1978). This particular package was chosen as it offers diverse robust professionally implemented numerical minimization functions and has been used previously on similar problems, for example, in showing that ML is inconsistent at identifying evolutionary trees when rates vary between sites (Lockhart et al. 1996). Specifically, the quasi-Newton method was used with quadratic forward estimates for initial minimization (usually accurate to 4-6 significant places), followed by application of the same quasi-Newton method, but using central quadratic estimates and a higher stringency (~7-9 places), followed by switching to a conjugate gradient method with central quadratic estimates to confirm a minimum was reached. The tree in the appendix was loaded to this optimizer using a PERL script written by PK.

The fossil calibration data includes that of the horse/rhino split within the order Perissodactyla. This is commonly taken to be the best data for calibrating any node of the tree deeper than 50 million years old and is estimated to have a standard error of ~1.5 million years (Waddell et al. 1999). The other calibration point used is that of human/tarsier within the order Primates. This calibration is strongly advocated by Beard (e.g., Beard et al. 1991) and is the best old calibration data within Primates, and perhaps the whole of the superorder Supraprimates (Waddell, Kishino, and Ota 2001) a clade also informally called Euarchontoglires (Waddell, Okada, and Hasegawa 1999). This calibration has an estimated mean and standard error of ~2.5 million years.

Maple version 10 or 11 was used to check algebraic expressions (Waterloo Maple 2007).

## 3 Results
### 3.1 Brownian motion along a tree

Kitazoe et al. recently described an important link between least squares estimation of relative node times for a weighted rooted evolutionary tree and a Brownian motion model (Kitazoe et al. 2007). (One-dimensional Brownian motion is described by a univariate Weiner mathematical process). They were specifically interested in a model where the log of the rate evolves by a Weiner process, as this was the exact form used by Thorne et al. (Thorne, Kishino, and Painter 1998). The key idea is to minimize the following sum of squares

$$SS_2 = \sum_{i=1}^{2t-4} \frac{(\ln(r_{anc_i}) - \ln(r_{des_i}))^2}{\sigma^2 \Delta t_i} = \frac{1}{\sigma^2} \sum_{i=1}^{2t-4} \frac{(\ln(r_{anc_i}) - \ln(r_{des_i}))^2}{\Delta t_i} \qquad \text{Eq (1)}$$

where $(\ln(r_{anc_i}) - \ln(r_{desc_i}))$ is the difference in rate of an ancestor/descendant pair of edges in the tree, $\sigma^2$ is the variance of the Brownian process, and $\Delta t_i = (t_{anc_i} - t_{des_i})/2$ or the time duration from the midpoint of the ancestral (*anc*) to the descendant (*des*) edge (further, the rates are the edge lengths of the tree divided by the difference in time of the node at each end of them). Penalizing the log of rates has the advantage of never having to deal with negative rates, and it implies proportional change from the previous state.

Treating $1/\sigma^2$ as a constant, and ignoring stochastic errors on edge length estimates, then minimizing this sum of squares will find ML-like solutions for the times if the process of rate change on the tree is a univariate Weiner process moving from the root to the tips. For such a model, the variance of the amount of rate change increases linearly with time. In this model, the estimated rate conceptually exists at the mid point in time of each edge (Kitazoe et al. 2007). Accordingly, the weighting term, $\Delta t$, is itself estimated as the difference in time from the mid-time of the ancestral edge to the mid-time of the descendant edge. This weight implies that a rate change of size *x* is much less surprising if it occurs after ten units of time rather than just one unit of time, for example.

### 3.2 Correlated Penalty Terms

Here we examine more closely that the rate is located at the mid-point of the ancestral edge and that two descendant edges issue below each ancestral edge in a binary rooted tree. Clearly, these sister descendant edges have potentially correlated rate penalty terms as shown in figure 1. After separation at the node, the two processes become independent, but they each include a component of time (and variance) from the ancestral edge. An improvement would be to minimize the double counting of correlated process, which is analogous to going from a weighted least squares to a generalized least squares solution.

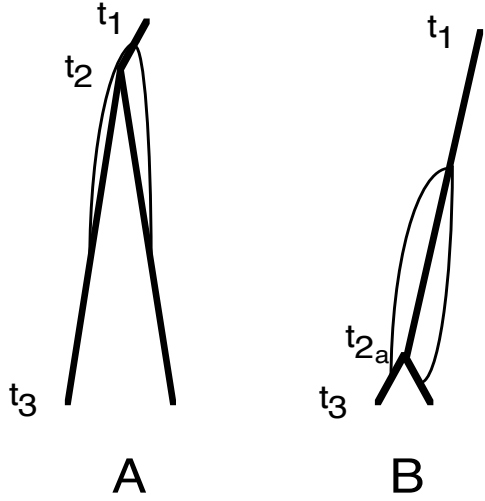

Figure 1. (A) Illustration of nearly uncorrelated rate change penalties and (B) nearly fully correlated rate change penalties. The arcs are of equal duration in time and trace the conceptual start and end points of the rate change penalty for the weighted least squares approximation to an ML solution for a Weiner (Brownian) process of rate change (Eq 1). The duration (in time) of the ancestral edge, $t_{anc}$, is $t_1$-$t_2$, and that of the descendant edges $t_{desc}$ = $t_2$-$t_3$. There is no penalty between the two descendant rates, yet they could vary in sign with respect to their difference from the ancestor.

From figure 1, it is also clear that equation 1 is not penalizing potentially large and unlikely changes in rate according to a Brownian model. For example, if the ancestral rate is 1 and the descendant rates are each ½, then equation 1 returns the same value as if one descendant is ½ and the other is 2. Moreover, if such a situation occurred on a branching pattern like that of figure 1b, it would imply a large change of rate (four times) in a very short period of time. One solution to this is to add a third penalty term, so,

$$SS_3 = Eq(1) + \frac{1}{\sigma^2} \sum_{i=1}^{2t-4} \frac{(\ln(r_{des_i}) - \ln(r_{des_j}))^2}{\Delta t_{ij}} \qquad \text{Eq (2)}$$

where $\Delta t_{ij}$ is one half of the duration in time of the two descendant lineages.

We might alternatively construct a generalized least squares (GLS) penalty of the rate differences about an internal node (excepting the root). Let **d** be a vector with components $d_1 = \ln(r_{anc}) - \ln(r_{des1})$, $d_2 = \ln(r_{anc}) - \ln(r_{des2})$, $d_3 = \ln(r_{des1}) - \ln(r_{des2}))$, then the sum of squares (SS) of the rate changes is obtained as,

$$SS_{GLS3} = \mathbf{d}^t \mathbf{V_x}^{-1} \mathbf{d} \qquad \text{Eq (3)}$$

where $t$ indicates transpose and $\mathbf{V_x}$ is the variance-covariance matrix of $\mathbf{x}$. Noting that the covariance of a univariate Weiner process from $t_0$ to $t_1$ and $t_0$ to $t_2$ is proportional to the lesser of $t_1$ and $t_2$, we arrive at

$$\mathbf{V_x} = \frac{\sigma^2}{2} \begin{bmatrix} t_{anc} + t_{des1} & t_{anc} & t_{des1} \\ t_{anc} & t_{anc} + t_{des2} & t_{des2} \\ t_{des1} & t_{des2} & t_{des1} + t_{des2} \end{bmatrix} \qquad \text{Eq (4).}$$

Ignoring, for the moment, the scalar $\sigma^2$, and after equating terms to single symbols for convenience, that is $d_1 = a$, $d_2 = b$, $d_3 = c$, $t_{anc} = x$, $t_{des1} = y$, $t_{des2} = z$, then

$$\mathbf{V}_\mathbf{x}^{-1} = \frac{1}{2xyz} \begin{bmatrix} xy+xz-zy & -xy-xz+zy & -xy+xz-zy \\ -xy-xz+zy & xy+xz+zy & xy-xz-zy \\ -xy+xz-zy & xy-xz-zy & xy+xz+zy \end{bmatrix} \qquad \text{Eq (5).}$$

After compiling and canceling terms we find that $\mathbf{x}^t \mathbf{V}_\mathbf{x}^{-1} \mathbf{x}$ simplifies to,

$$SS_{GLS3} = \frac{1}{2xyz} \begin{bmatrix} (xy+xz+yz)((a-b)^2+(a-c)^2+(b-c)^2) - 2\{(a-b)(a-c)(xy+xz-yz) \\ +(a-b)(b-c)(xy-xz+yz)+(a-c)(b-c)(-xy+xz+yz)\} \end{bmatrix}$$

Eq (6).

However, the reported sum of squares changes if we take $d_3 = \ln(r_{des2}) - \ln(r_{des1})$, which is undesirable. An average of both ways is one possible solution.

If we limit ourselves to penalizing just (a-b) and (a-c), then taking into account variances and covariances, the corresponding sum of squares is equal to,

$$SS_{GLS2} = \frac{2}{xy+xz+yz}\left[(a-b)^2(x+z)+(a-c)^2(x+y)-2(a-b)(a-c)x\right] \qquad \text{Eq (7).}$$

If we set the covariance terms in $\mathbf{V}$ to zero, $SS_{GLS2} = SS_2$. As with Eq(1) or (2), these terms are updated on each cycle of the optimization of divergence times.

**3.3 Root Penalty Terms**

The previous terms may be used for all ancestor/descendant pairs in the tree, but the root requires special consideration. It is possible to ignore any root terms completely, for example (Kitazoe et al. 2007). An alternative option is to add another free parameter to the model, that is $r_0$, a rate for the root, and optimize this along with the times. This does not present a significant problem, since the value of $r_0$ that imposes least penalty will be a weighted (by the inverse of time) average of the two descendants of the root. Thus, the addition of a free root rate is acting more like a constraint than a free parameter. Since the root rate is at the root node, there is no need to reweight for correlated penalties of the root descendants. The rate at the root therefore seems identifiable, but estimating the duration and rate of an edge above the root is not identifiable.

The model corresponding to unweighted least squares, which is like the penalty proposed by Sanderson, is appropriate if rates undergo a random walk on the tree (Kitazoe et al. 2007). The two descendants of the root may be considered independent draws from a normal distribution of the same variance as the rest of the tree. In this case, the variance of their difference is twice that of a single draw going from one ancestor to one descendant. Therefore, the appropriate penalty for these two terms is $\frac{(r_{des1}-r_{des2})^2}{2}$.

Note, Kitazoe et al. (2007) also present a model where the weighting factors are the branch lengths. This would correspond to a Weiner process with respect to edge length or evolutionary divergence. In this model, the time terms are replaced by edge weights (which are time multiplied by substitution rate). This is an intriguing model since it suggests that the variance of rate change will increase not just with time, but also with increasing rates of evolution. Thus if using a log rate form, it suggests that the percentage change in rate in the mouse lineage is expected to be larger than the percentage change of rate in the human lineage (which has an average rate of ~1/3 that of the mouse).

### 3.4 Adjusting for unknown variance

The penalty term in equation (1) initially looks like it can be minimized while effectively ignoring the term $1/\sigma^2$. That is like using ordinary least squares (OLS) fitting of trees to a distance matrix (Swofford et al. 1996), However, $1/\sigma^2$ is not a constant, since unlike its counterpart in OLS estimation of a tree from distances, it itself changes with the times which are being simultaneously estimated. That is, if all the times are doubled, then $\sigma^2$ will reduce by a factor of 2 and the sum of squares, ignoring $1/\sigma^2$, will reduce by a factor of 2, all else remaining equal. This is the same type of problem encountered in Waddell, Kishino, and Ota (2007) where the sums of squares of for trees to a distance matrix need to be normalized with regard to the scale of the distance matrix if they are to provide a general measure of goodness of fit comparable across distance matrices. Thus, the scale effect of time in equation (1) is analogous to the scale effect of distance considered in Waddell, Kishino and Ota (2007), e.g., equation A3.

If the root is constrained to a constant, for example by a fossil calibration, this effect is largely negated. However, if only times below the root are constrained, then this ameliorates the effect but does not eliminate it; the bias is for the penalty terms deeper in the tree to become lower by simply increasing the time intervals between nodes.

One way to compensate for this tendency is to penalize the weighted sum of squares by the ratio of the sum of all the time differences used to weight the sum of squares relative to that of a previous solution. If there is an obvious local minima to the sum of squares achieved without this term, then this may be used for the purpose of defining the denominator against which other solutions are assessed and penalized. Note that this additional "penalty term" will reduce the adjusted sum of squares when sum of times becomes less. This additional term will also have the effect of making the confidence interval for the root estimated by fixing this value and reestimating all other values more symmetric.

There is a similar problem with minimizing the square of the difference of rates (abbreviated $R^2$), that is, Eq (8)

$$SS_{R^2} = \frac{1}{\sigma^2}\sum_{i=1}^{2t-4}(r_{anc_i} - r_{des_i})^2 = \frac{1}{\sigma^2}\sum_{i=1}^{2t-4}(e.l._{anc_i}/\Delta t_{anc_i} - e.l._{des_i}/\Delta t_{des_i})^2 = \frac{4}{\sigma^2}\sum_{i=1}^{2t-4}(e.l._{anc_i}/(2\Delta t_{anc_i}) - e.l._{des_i}/(2\Delta t_{des_i}))^2$$

Thus, if the term $1/\sigma^2$ is ignored, there should still be compensation for the magnitude of time and here it is even worse than for equation (1). The sum of squares without term $1/\sigma^2$ is decreasing, not with the sum of the differences of times, but with the square of the sum of the differences of times. If we take logs of rates in the former equation the effect disappears since doubling of time does not change the expected sum of squares. Worse still, if straight rates and a weighting by time are combined (rates undergoing Brownian motion), then doubling the time, all else held constant, results in the sum of squares decreasing by a factor of eight. Alternatively, weighting by branch length is a constant that does not change with time. Finally, using the inverse of rate in place of the rate, effectively puts time on the top, so that a doubling of time increases the sum of squares by a factor of four due to this factor alone. Taking logs of the inverse rate and squaring the difference removes the scale effect and also results in exactly the same penalty as the log of the rates. For these reasons, log of the rate seems to be something of a "most natural" model.

**3.5 A weighted-least squares framework including fossil data**

Most researchers have treated fossil data that allow an estimate of the age of a node in the tree as either calibration point without error, or as a pair of upper and lower constraints (Thorne, Kishino, and Painter 1998). In contrast, our own work has specified them as a mean and standard deviation (Waddell and Penny 1996, Waddell et al. 1999). There, assuming that the total error on a calibrated node is made up of errors from the fossils and independent errors on edge lengths of the tree, a confidence interval on the unknown time at the end of the ratio of edge lengths is made. Here we combine this weighted least squares approach with the use of least squares penalties of the rate of molecular evolution.

The calibration point we use is that of the horse-rhino divergence (Waddell et al. 1999), with a mean of 55 million years and a standard deviation of 1.5 million years. Its weighted least squares contribution to the total sum of squares is equal to $(x-55)^2/2.25$, where $x$ is the assumed age of the node.

In order to combine a fossil least squares term with the residual of the evolution of evolutionary rate, we need to estimate $\sigma^2$. There are $2N-2$ pieces of information (the edge lengths), and $N-1$ parameters are estimated. Thus, under the model, the residual sum of squares ($SS_3$) should be distributed as chi-square with degrees of freedom $N-1$. Therefore $\frac{SS}{\sigma^2} = N-1$, $\frac{SS}{N-1} = \sigma^2$. In a case considered later $N$ is 32, $SS_3 = 0.1375$, so $\sigma^2 = 0.004435$, and the standard deviation of the evolution of the evolutionary the rate is 0.06659 per million years. Fluctuations in the node times due to ancestral polymorphism or other causes, are expected to increase the sum of squares and hence also increase the estimate of $\sigma^2$. This is a desirable robustness property (Kitazoe et al. 2007).

Assuming errors on fossils are independent of fluctuations in the rate of molecular evolution, the total residual is the normalized term $SS_3$, plus the fossil term (the $SS_3$ term retains the appropriate inflation term from section 3.4). In effect, the fossil calibrated node is contributing two penalties; one with respect to change of rate, the other with respect to agreement with fossil data. If there is just one fossil calibration point, then the contribution due to the fossil is minimized to zero by taking the age of this node as the mean age. If there is a second fossil calibration point added, and errors on it are considered to be normally distributed with known mean and variance, then its weighted least squares term is added in to the total sum of squares. If the fossil times cannot both fit perfectly on the tree, which is expected, then all node times in the tree must be reoptimized to minimize the total weighted sum of squares. If the fossils agree with each other and the estimated node times expect for expected error, the total sum of squares should increase by 1 by adding fossil data for a second node. A 95% confidence interval on how much the total sum of squares should increase by is approximately 0 to 3.84 (based on a $\chi^2$ distribution with 1 degree of freedom). If it exceeds this, there is clear evidence that the two fossil calibration points and the relative molecular times do not agree. The error could be in either or both fossil estimates of node times. It could also be a failure to accurately estimate the relative node times from the tree, due to the true process of rate evolution not following Brownian motion of the log of the rate. It could even be a failure to estimate the true relative edge lengths of the input tree accurately, which with increasingly long sequences is increasingly due to systematic error, e.g. due to an overly simple approximation to the evolutionary process of character change.

This approach also opens a direct avenue to bootstrapping the fossil data on the ages of nodes. This is achieved by drawing a bootstrap sample, where each fossil calibration "point" is given an equal probability, and the bootstrap weight is transferred to the sum of square term associated with that fossil information. With two points the reweighting vectors and their relative probabilities are, respectively, [2,0], p=0.25, [1,1], p=0.5, and [0,2], p=0.25. That is, there is a 25% chance that the first fossil calibration point will be given double weight and the second ignored, a 50% chance both will be included with a relative weight of 1, and a 25% chance the first calibration point will be ignored and the second point given double weight. The resulting total interval is built over both of these, and it could be much wider than using either point alone if the two points disagree strongly.

**3.6 Reestimated divergence times of placental mammals**

The first assessments we make are on the age of different clades using different penalty functions, but the same data. In all these cases (table 1) a single fixed calibration of horse/rhino at 55 mybp is used. The first thing to notice in table 1 is that the age of all unconstrained nodes in the tree become abnormally old if the function $R^2$ (Eq 8) is used (second column), and the sum of squares is heading towards zero (but can never get there due to the constraint, c.f. Kitazoe et al. 2007). Adding in a penalty for the difference in rates on either side of the root, which is equivalent to estimating a rate at the root, is a small term that makes very little difference in this case (third column). This behavior seems to be a direct consequence of not using a scale factor; use of a scale factor gives a root value similar to the other methods in table 1 (results not shown).

The second penalty used in table one is that of $SS_2$ (Eq 1), in the same form as that used in Kitazoe et al. (2007). It gives slightly older ages for the root on mammals than that example, which is not surprising given the differences in the tree used, and that this example uses a single calibration point. Adding in an estimated rate at the root and its associated penalty increases the sum of squares by a small amount, and again, causes little difference to the estimated ages of mammal clades.

Taking into account the correlation along the shared path of an ancestor ($GLS_2$, Eq 7)

makes relatively little difference to the inferred ages, nor does adding in its root term. The use of $SS_3$ results in ages becoming slightly older. It also results in a notable increase in the sum of squares, probably because it is penalizing changes of rate between descendants, which can be marked in some cases. Here too the addition of the root term has a minimal effect. Looking at rows 4 and 5 of table 1, we see that the rates of the two descendant lineages are fairly similar, even when they are not forced to be. If this changes, e.g. due to the form of the tree, with a uniformly slow group on one side and a uniformly fast group on the other, then one can imagine the addition of a root rate having a greater effect. In all cases the free root rate was roughly the unweighted average of the two descendants, but this could of course change with the Brownian models if one of the descendant edges persisted many times longer than the other.

Column one of table 1 considers the coefficient of variation between divergence times due to the use of different forms of the Brownian penalty (it ignores the times in columns two and three and is expressed as a percentage). There is no clear trend of the nodes close to the calibration point showing less than the average amount of variability. A surprise is that while the root may show the largest absolute fluctuation, it is relatively recent events (e.g., the mouse/rat and Cetacea splits) that are most labile.

Table 1. Comparison of estimated divergence times of placental mammals using the horse/rhino split at 55 mybp and eight different penalties. The penalties are, difference of rates squared, $R^2$, difference of log rates squared weighted by time, $\ln R/T$, difference of log rates squared weighted by time taking into account correlation from ancestor, $GLS_2$, and log of rates squared weighted by time including a penalty between descendants, $SS_3$, each with or without a root penalty (if with "+R"). No scale adjustments were used. Names of clades follow Waddell, Kishino and Ota (2001) and references therein.

| Type | $R^2$ | $R^2$ +R | $\ln R/T$ | $\ln R/T$ +R | $GLS_2$ | $GLS_2$ +R | $SS_3$ | $SS_3$ +R | Clade |
|---|---|---|---|---|---|---|---|---|---|
| SS | 7.50E-04 | 7.50E-04 | 62.499 | 62.502 | 86.311 | 86.312 | 137.472 | 137.506 | |
| Rates[a] | 7.91E-05 | | 1.527 | | 1.509 | | 1.457 | | |
| cv[b] | 5.39E-05 | 6.65E-05 | 1.508 | 1.519 | 1.516 | 1.512 | 1.512 | 1.482 | |
| 0.4 | 3.21E+06 | 3.22E+06 | 124.4 | 124.5 | 124.7 | 124.7 | 125.5 | 125.4 | Placentalia/Root |
| 0.3 | 3.10E+06 | 3.06E+06 | 118.5 | 118.6 | 118.7 | 118.7 | 119.4 | 119.2 | Exafricomammalia |
| 0.4 | 2.64E+06 | 2.64E+06 | 103.9 | 104.1 | 104.3 | 104.2 | 105.1 | 104.7 | Afrotheria |
| 0.1 | 2.93E+06 | 2.95E+06 | 108.2 | 108.2 | 108.2 | 108.2 | 108.5 | 108.4 | Boreotheria |
| 0.4 | 2.46E+06 | 2.46E+06 | 96.2 | 96.4 | 96.6 | 96.5 | 97.4 | 96.9 | Afroinsectiphilla |
| 0.0 | 2.59E+06 | 2.59E+06 | 94.9 | 94.9 | 94.9 | 94.9 | 95.0 | 95.0 | Laurasiatheria |
| 0.1 | 2.81E+06 | 2.83E+06 | 100.2 | 100.2 | 100.0 | 100.0 | 100.1 | 100.0 | Supraprimates |
| 0.1 | 2.41E+06 | 2.41E+06 | 87.8 | 87.8 | 88.0 | 88.0 | 88.0 | 88.0 | Scrotifera |
| 0.7 | 2.25E+06 | 2.25E+06 | 80.4 | 80.4 | 79.6 | 79.6 | 79.2 | 79.2 | Eulipotyphla |
| 0.1 | 2.77E+06 | 2.77E+06 | 98.1 | 98.1 | 97.9 | 97.9 | 97.9 | 97.9 | Euarchonta |
| 0.2 | 2.63E+06 | 2.63E+06 | 93.5 | 93.5 | 93.2 | 93.2 | 93.0 | 93.0 | Glires |
| 0.2 | 4.68E+03 | 4.68E+03 | 82.6 | 82.6 | 83.0 | 83.0 | 82.9 | 82.8 | Fereuungulata |
| 0.2 | 2.07E+06 | 2.07E+06 | 67.7 | 67.7 | 67.8 | 67.8 | 67.6 | 67.6 | Chiroptera |

| | | | | | | | | |
|---|---|---|---|---|---|---|---|---|
| 0.3 | 2.34E+06 | 2.34E+06 | 88.6 | 88.6 | 88.3 | 88.3 | 88.1 | 88.1 | Primates |
| 0.3 | 1.76E+06 | 1.76E+06 | 63.8 | 63.8 | 63.9 | 63.9 | 63.5 | 63.5 | Lagomorpha |
| 0.6 | 2.33E+06 | 2.33E+06 | 84.5 | 84.5 | 84.0 | 84.0 | 83.4 | 83.4 | Rodentia |
| 0.2 | 4.31E+03 | 4.31E+03 | 77.7 | 77.7 | 78.1 | 78.1 | 77.9 | 77.9 | Ferae |
| 0.3 | 3.42E+02 | 3.42E+02 | 78.5 | 78.5 | 79.0 | 79.0 | 78.8 | 78.7 | Euungulata |
| 1.9 | 1.24E+06 | 1.24E+06 | 29.8 | 29.8 | 31.0 | 31.0 | 30.8 | 30.8 | megabats |
| 0.3 | 2.24E+06 | 2.24E+06 | 86.0 | 86.0 | 85.7 | 85.7 | 85.4 | 85.4 | human/tarsier |
| 1.2 | 2.08E+06 | 2.08E+06 | 72.9 | 72.9 | 71.9 | 71.9 | 71.0 | 71.0 | hystrich./murid_ |
| 0.5 | 2.98E+03 | 2.98E+03 | 55.8 | 55.8 | 56.5 | 56.5 | 56.2 | 56.2 | Carnivora |
| **0.0** | **5.50E+01** | **5.50E+01** | **55.0** | **55.0** | **55.0** | **55.0** | **55.0** | **55.0** | **Perissodactyla** |
| 0.9 | 2.51E+02 | 2.51E+02 | 60.6 | 60.6 | 60.9 | 60.9 | 59.7 | 59.7 | Cetartiodactyla |
| 1.4 | 9.68E+05 | 9.68E+05 | 50.7 | 50.7 | 49.9 | 49.9 | 49.1 | 49.1 | Anthropodea |
| 2.5 | 4.26E+05 | 4.26E+05 | 20.4 | 20.4 | 19.4 | 19.4 | 19.4 | 19.4 | mouse/rat |
| 0.4 | 2.42E+03 | 2.42E+03 | 45.5 | 45.5 | 46.0 | 46.0 | 45.7 | 45.7 | Caniformia |
| 1.0 | 2.28E+02 | 2.28E+02 | 55.3 | 55.3 | 55.5 | 55.5 | 54.4 | 54.4 | Artiofabula |
| 1.1 | 1.96E+02 | 1.96E+02 | 47.2 | 47.2 | 47.4 | 47.4 | 46.3 | 46.3 | Cetruminantia |
| 1.1 | 1.74E+02 | 1.74E+02 | 41.9 | 41.9 | 42.0 | 42.0 | 41.1 | 41.1 | Whippomorpha |
| 2.0 | 8.25E+01 | 8.25E+01 | 18.6 | 18.6 | 19.3 | 19.3 | 18.5 | 18.5 | Cetacea |
| $X^{2\,c}$ | 4.53E+07 | 4.53E+07 | 0.2 | 0.2 | 0.1 | 0.1 | 0.0 | - | |

[a] The rate on either side of the root, or the rate at the root. These numbers and the $SS$ are × 1000.

[b] The coefficient of variation (standard deviation/mean) × 100% of the times in a row for all but the $R^2$ penalties.

[c] An $X^2$ statistic comparing each column of estimated ages with those of the $SS_3$+R penalty.

Figure 2 shows that the impact of the scale factor. Without it there is clearly a second minimum. In this case, with $SS_3$ and a root penalty, without the scale factor the global minimum was not reached until the root was older than the earth! Incorporating the scale factor shifts the ages of mammal groups to yet younger ages; in this case the root to 112.5 million years and there was no sign of a second minimum. Its use also results in a much more symmetric minimum, although the confidence interval (CI) is still skewed towards older times. Allowance for the uncertainty of the single fossil calibration point increases the width of the CI by about 10%. As explained above, the procedure for estimating the CI is to firstly find the times that minimize the penalty $SS_3$ +R allowing for the scale effect and with the horse/rhino node fixed at 55 mybp. The scale effect penalty is equal to the sum of all times used in the denominator at the current assessment divided by the sum of all times at an earlier iteration. When this minimal sum of squares is found, then $\sigma^2$ is estimated and fixed. The sum of squares due to the model of rate change is set to the degrees of freedom of the system. This is its expected value if the model is correct; if it is not correct it is expected to underestimate the true sum of squares. To this sum of squares is added the mean age of the calibration node minus its assumed value all squared, divided by the variance of this point (2.25 in this case). This does not alter the minima, but it does increase the CI as altering the exact age of the calibration node is now possible and if by a relatively small amount, this will add only a small penalty. The CI is much wider than that due to the uncertainty in the calibration point due to the uncertainty in the assumed Brownian model of rate evolution.

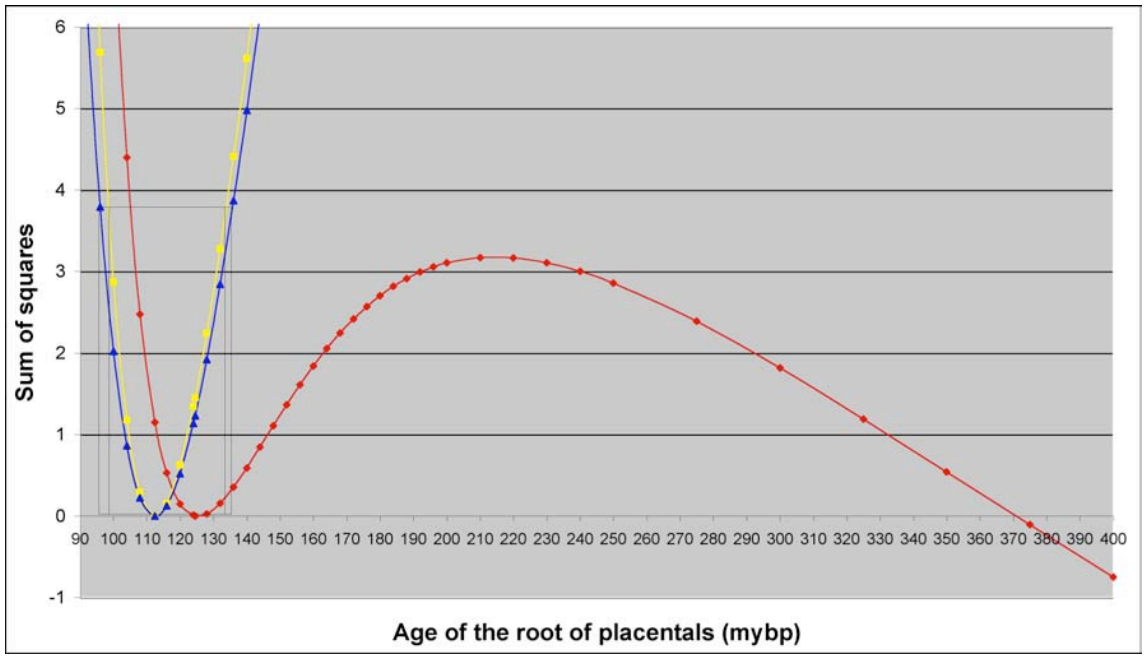

Figure 2. Curves of the least squares fit obtained by setting the root to a specific value and reoptimizing all other parameters (times, root rate) except for the calibration point of horse/rhino at 55 million years. In all cases the rate change penalty function is $SS_3 + R$. The red curve does not include a scale term and two minima are apparent. The blue line, which includes a scale term, is the same as the yellow line except it measures the total sum of squares including that due to the fossil calibration data. The 95% confidence interval is indicated by the gray lines and occurs when the normalized (to equal the degrees of freedom) of the sum of squares has deteriorated by 3.84 from its minimum. Uncertainty from the single fossil calibration increases the confidence interval of the age of the root by about 10% in this case.

Figure 3 (the blue curve) shows the effect of adding a second calibration point distant to the first one. In this case it is the estimated age of the human/tarsier split. It sharply reduces the estimated age of the root to 94.2 million years. Also, its addition strongly inflates the sum of squares, which go from 31 with just one calibration point to a minimum of 49.0. The components of this are 40.8 from the Brownian model of rate change, 2.7 from the horse/rhino node being forced lower and 5.5 from the human tarsier node being pulled higher. If everything was correct and followed assumptions, then a 95% CI on the size of the increase due to fixing another node should have been 3.84; the actual increase was far larger and suggests that something is amiss. It could be an error on the estimated age or variability of either fossil calibration, it could be that our assumptions about how rates change are wrong, or it could be both things. It could also be the tree "data." In this case the weighted tree enters calculations effectively as data, but either the topology or the edge lengths (or both) could be non-negligibly wrong.

The curve in figure 3 is asymmetric and this results in asymmetric confidence intervals on the age of the root. Interestingly, the confidence intervals do admit an age of the root (85 mybp) that is young enough to satisfy many of those paleontologists that see no evidence for modern placental mammal groups in the Cretaceous. The right hand side of the curve keeps increasing, showing no signs of reaching a constant value or decreasing for ages extended out to the age of the earth (about 4.6 billion years).

Figure 3 also shows a red line corresponding to the normalized sum of squares when a human / tarsier calibration alone is used. This calibration, as mentioned earlier, is one of the most reliable of all the deeper calibrations in the primate part of the tree. Using the human / tarsier calibration alone, a 95% CI is ~71.2 to 90.7, while a 99 % CI is ~68.6-94.6.

Table 2 shows the estimated ages of all the groups on the tree using both calibration points. Many of these ages seem quite or even very reasonable with regard to the overall fossil record. The most problematic are the last four ages which all fall within the Cetartiodactyla. These are all too young by about 15 million years. At least two of these nodes have good fossil data that suggest the estimates here are too young (Cetacea and Whippomorpha). The exact reason is unclear but there are some considerable rate changes in this part of the tree. If the ancestral edge leading to Cetartiodactyla is either over estimated in length, or incurred a marked acceleration then deceleration, this might help to explain some of the discrepancy.

A retort might well be that the problem is with the human / tarsier calibration being too young. However, the Cetartiodactyla dates are still too young by about 15 million years if the only calibration used is that of horse/rhino, and as already mentioned, this is by far the best calibration in the tree. Indeed if the age of Whippomorpha is constrained to an age of 50 mybp, only the horse/rhino calibration is used and everything else is allowed to reoptimize, then the total sum of squares still increases by 8.8, which indicates a real clash. This is consistent with the claims of Waddell, Kishino and Ota (2001) that these major calibration points of the placental tree give rise to very different ages of the root and significantly clash with each other.

An age of 50 mybp is an absolute minimum age for the Whippomorpha node, since the latest claims for the oldest whale is *Himalayecetus subathuensis* at 53.5 mybp, with many other potential whale species in the 5 million years after this. However, even with this constraint, the age of Cetacea is still only 22.9 mybp, which remains problematic if *Llanocetus denticrenatus* (at 34. Mybp, Fordyce 1989) is indeed more closely related to living baleen whales than living toothed whales. What is also important to realize is that the horse/rhino calibration is one of the few calibration points that seems to be bounded above; this is because there is evidence of what seem to be nearly direct ancestors and there is also another well traced split (rhino/tapir) soon after (Waddell et al. 1999). If many fossil constraints are used with lax upper limits and tight lower limits, then a trend is for the root to be pushed older, simply because there is an argmax effect occurring with regard to the lower bounds. This can become very pronounced if there is a wide fluctuation of the estimated relative molecular times about their true values.

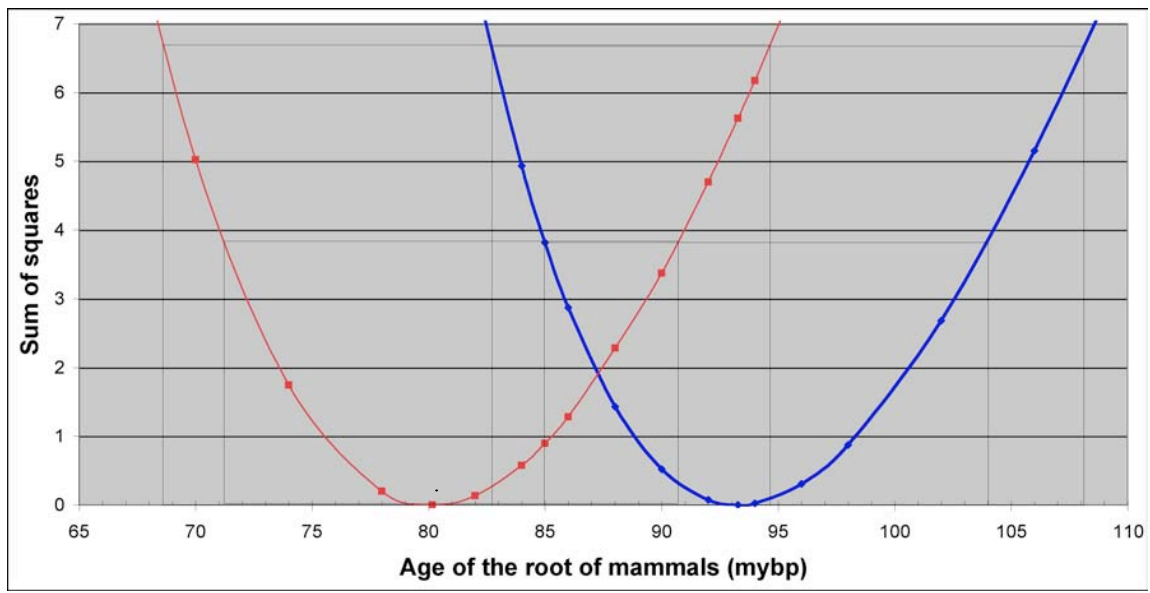

Figure 3. Curve of the total least squares fit (normalized $SS_3$ plus fossil calibration terms) obtained by setting the root to a specific value and reoptimizing all other free parameters (times, root rate, calibration points). The calibration points are horse/rhino at 55 mybp (s.e. 1.5 myr) and human/tarsier also at 55 mybp (s.e. 2.5 myr) for the blue curve, and just the human/tarsier calibration for the red curve. Shown are a 95% (critical value 3.84) and a 99% (critical value 6.63) confidence interval on the age of the root. More exact confidence intervals (CI), estimated on fine scale graphs of the interpolated curves, were 85.0 to 104.0 for a 95% CI, and 82.7 to 108.1 for a 99% CI for the red curve.

Table 2. Divergence times of placental groups minimizing a total least squares approach. The penalty for rate change was $SS_3$ including a root penalty, while the calibration points were horse/rhino and human/tarsier.

| Clade | Age [a] | Clade | Age | Clade | Age | Clade | Age |
|---|---|---|---|---|---|---|---|
| Placentalia/Root | 93.3 | Eulipotyphla | 62.2 | Ferae | 62.3 | Anthropodea | 33.7 |
| Exafricomammalia | 88.8 | Euarchonta | 72.4 | Euungulata | 63.0 | mouse/rat | 14.1 |
| Afrotheria | 78 | Glires | 68.5 | megabats | 25.8 | Caniformia | 38.2 |
| Boreotheria | 81.1 | Fereuungulata | 65.3 | human/tarsier | **60.9** | Artiofabula | 46.5 |
| Afroinsectiphilla | 72.2 | Chiroptera | 55.0 | hystrich./murid_ | 51.7 | Cetruminantia | 39.9 |
| Laurasiatheria | 72.5 | Primates | 63.2 | Carnivora | 46.6 | Whippomorpha | 35.5 |
| Supraprimates | 74.3 | Lagomorpha | 45.8 | Perissodactyla | **52.6** | Cetacea | 16.0 |
| Scrotifera | 68.3 | Rodentia | 61.0 | Cetartiodactyla | 50.7 | | |

[a] In mybp.

## 4 Discussion

The analysis in figure 3 suggests that young ages for the root of placental mammals are still possible, even if the model of log rate evolution was Brownian. It should also be emphasized that like Kitazoe et al. (2007) the confidence interval takes into account variability in edge lengths of the tree due to ancestral polymorphism, but it does not take into account bias in the age of the root (and other nodes) due to this cause. If the fossil times are bootstrapped, then ¼ of the time the fossil data will all come from the human/tarsier split. As seen in Waddell, Ota and Kishino (2001), use of this calibration point results in a root that is as young as 80 million years. It is, in principal, very important to bootstrap the fossil calibrated nodes, since it is very happenstance where these occur, and their quality also varies widely (so much so that biologists generally ignore the fossil data relevant to the age of the vast majority of nodes in the tree).

The addition of a scale term is seen to be very important since we do not know the variance of the Brownian process separate to time. Without it many of the penalty functions return far smaller sums of squares by simply doubling times. In many cases, constraining one or more nodes lower in the tree is insufficient to prevent the penalty becoming smaller as times near the root are increased. In Waddell, Ota and Kishino (2001) the scale effect was largely controlled by fixing the root to an arbitrary value of one then solving for all other times. Only then was the specific calibration applied by multiplying all nodes in the tree by the calibration point's age divided by its relative molecular time. While this is useful, the current strategy for estimating and using a scale parameter should be more accurate.

For these analyses we assumed the tree used by Waddell, Kishino and Ota (2001). Recently, phylogenomic data has suggested that the very similar tree of Waddell, Okada and Hasegawa (1999) was more accurate in some details. For example, rather than putting Afrotheria first, placing Afrotheria sister to Xenathra in the clade Atlantogenata (Waddell et al. 1999). Support for this arrangement comes from analyses of LINE sequences and highly conservative indels (Waddell et al. 2006, Waters et al. 2007). As Kitazoe et al. (2007) point out, this rearrangement makes a relatively small difference to the age of the root. However, with other the other types of rearrangement expected in the tree (indicated by arrows in Waddell, Okada and Hasegawa 1999 and the figure of Waddell, Kishino and Ota 2001), the impact of each of these needs to be checked.

Overall, the analyses suggest that the divergence times of placental mammals are much less certainly inferred from molecular data than is commonly assumed. It is also important to identify apparent contradictions between fossil calibrations. Here, human/tarsier, horse/rhino and whale/hippo calibrations all seem to clash to a significant degree. However tempting, this should not automatically lead to censorship (e.g. Near and Sanderson 2004) of "unreliable" fossil calibration data. *A priori*, the problem could equally lie with the reconstruction of relative times based on the molecular data (e.g., Kitazoe et al. 2007), in which case all fossil times may be valuable. Significantly, systematic errors in edge length estimation will directly affect estimated times of nearby nodes. This is somewhat in contrast to the fossil data where at least some major errors are best conceived of as independent between data used for calibrating different nodes. It is temping to believe that problems with the molecular data will disappear by using many long genes. However, an ongoing concern is that discussed in Waddell et al. (2001). That is, the main autocorrelated rate effect is shared by all genes in a genome, and on top of this each gene has largely unpredictable rate changes due to natural selection. Thus there may only be a single underlying autocorrelated process, with noise added on top of it. The combination of many genes will reduce the variance due to this noise, and if the noise is unbiased, that contribution will diminish with more genes being added. However, there is still only a single core autocorrelated process and thus limited scope for eliminating the uncertainty due to a stochastic process of rate evolution.

Future directions include adding a term to combat rates of evolution tending to rise towards the tips of the tree (or, conversely, be too low towards the root). This is due to the log rate not being able to take on negative values, so with time it is expected to increase in value. A term like this is used within the program multidiv time.

## Acknowledgements

This work was supported by NIH grant 5R01LM008626 to PJW. Thanks to Hirohisa Kishino and Yasu Kitazoe for helpful general discussions and Phillip Buckhaults for assistance with formatting in Word.

## Author contributions

PJW originated the research, developed methods, gathered data, ran analyses and prepared figures and wrote manuscript. PK working as a student assistant wrote a PERL script ("The_Times") to parse trees to the solver and assisted in analyses.

## Appendix 1

The weighted tree used as the basis of divergence time estimates. The rooting point of the placental mammals is estimated by where the two marsupial outgroups (Kangaroo and Opossum) join the tree. Thus, it is the rooted subtree of placentals that is the data used in the analyses of divergence time.

((((((((Pangolin:0.12351300,((Dog:0.05662000,Seal:0.05627600):0.01275100,Cat:0.06044500):0.02683000):0.00633800,((Horse:0.04873800,Rhino:0.05013000):0.02220300,(((Cow:0.08321700,((Baleen_whale:0.02709900,Sperm_whale:0.05253800):0.04539000,Hippo:0.06537400):0.00931700):0.01348200,Pig:0.08450400):0.00823800,Lama:0.08242800):0.02519900):0.00508000):0.00669600,((Megabat:0.04427000,Flying_fox:0.03066600):0.04823500,Jamacian_fruit_bat:0.13046300):0.02891700):0.00985200,(Hedgehog:0.26972200,Shrew/Mole:0.10455400):0.02510100):0.02013600,(((((Human:0.09177400,Old_world_monkey:0.13586800):0.07007900,Tarsier:0.11588500):0.00442500,Galago:0.14388600):0.01596200,Tree_shrew:0.15249100):0.00347700,((Pika:0.10222900,Rabbit:0.08866300):0.04695200,((Guinea_pig:0.16552200,(Mouse:0.05956000,Rat:0.06051900):0.14675500):0.02376900,Squirrel:0.11229900):0.01615700):0.01139500):0.01315000):0.01618500,Armadillo:0.15230800):0.00900200,(Elephant:0.15464500,(Aardvark:0.09320700,Tenrec:0.21955200):0.01155700):0.03096500):0.06466392,(Kangaroo:0.09674700,Opossum:0.10290800):0.21183008);